\documentclass[journal=jpccck,manuscript=letter,layout=twocolumn]{achemso}

\usepackage[version=3]{mhchem}
\usepackage{bm}
\usepackage{physics}
\usepackage{graphicx}
\graphicspath{ {./figures/} }
\usepackage{caption}
\usepackage{float}
\usepackage{booktabs}
\usepackage{multirow}
\usepackage{subcaption}
\usepackage{xcolor}
\usepackage{hyperref}
\hypersetup{pdfborder=0 0 0,colorlinks=true,citecolor=blue,linkcolor=blue}

\SectionNumbersOn

\title{Dynamic Distortions of Quasi-2D Ruddlesden-Popper Perovskites at Elevated Temperatures: Influence on Thermal and Electronic Properties}

\author{Raisa-Ioana Biega}
\email{r.i.biega@utwente.nl}
\affiliation[UTwente]
{MESA+ Institute for Nanotechnology, University of Twente, 7500 AE Enschede, The Netherlands}
\alsoaffiliation[Physics Uni Bayreuth]
{Department of Physics, University of Bayreuth, 95440 Bayreuth, Germany}

\author{Menno Bokdam}
\affiliation[UTwente]
{MESA+ Institute for Nanotechnology, University of Twente, 7500 AE Enschede, The Netherlands}

\author{Kai Herrmann}
\affiliation[Chemistry Uni Bayreuth]
{Department of Chemistry, University of Bayreuth, 95440 Bayreuth, Germany}

\author{John Mohanraj}
\affiliation[MC I]
{Applied Functional Polymers, University of Bayreuth, 95440 Bayreuth, Germany}

\author{Dominik Skrybeck}
\affiliation[MC I]
{Applied Functional Polymers, University of Bayreuth, 95440 Bayreuth, Germany}

\author{Mukundan Thelakkat}
\affiliation[MC I]
{Applied Functional Polymers, University of Bayreuth, 95440 Bayreuth, Germany}
\alsoaffiliation[BPI]
{Bavarian Polymer Institute, University of Bayreuth, 95440 Bayreuth, Germany}

\author{Markus Retsch}
\affiliation[Chemistry Uni Bayreuth]
{Department of Chemistry, University of Bayreuth, 95440 Bayreuth, Germany}

\author{Linn Leppert}
\email{l.leppert@utwente.nl}
\affiliation[UTwente]
{MESA+ Institute for Nanotechnology, University of Twente, 7500 AE Enschede, The Netherlands}

\begin{document}

\begin{abstract}\label{abstract}
Ruddlesden-Popper hybrid halide perovskites are quasi-two-dimensional materials with a layered structure and structural dynamics that are determined by the interplay between the organic and inorganic layers. While their optical properties are governed by confinement effects, the atomistic origin of thermal and electronic properties of these materials is yet to be fully established. Here we combine computational and experimental techniques to study A$_2$PbI$_4$ (A=butylammonium (BA), phenethylammonium (PEA)) Ruddlesden-Popper perovskites and compare them with the quintessential perovskite CH$_3$NH$_3$PbI$_3$. We use first-principles density functional theory, molecular dynamics simulations based on machine-learned interatomic potentials, thermal measurements, temperature-dependent Raman spectroscopy, and ultraviolet photoelectron spectroscopy, to probe the thermal and electronic properties of these materials at elevated temperatures. Our molecular dynamics simulations demonstrate that dynamic fluctuations in the organic sublattice determine the bulk-average distortions of these materials at room-temperature, explaining significant differences in their electronic density of states close to the Fermi level. Furthermore, by analysing the organic layer dynamics in BA$_2$PbI$_4$ we provide a mechanistic explanation for the phase transition of this material at 274\,K and observations from Raman measurements. Our results highlight the role of the organic interlayer for the electronic and thermal transport properties of Ruddlesden-Popper perovskites, paving the way for the design of new hybrid materials for tailored applications.

\begin{tocentry}
    \centering
    \includegraphics{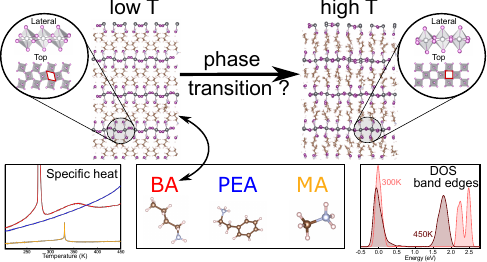}
\end{tocentry}

\end{abstract}

\newpage

\section{Introduction}
Hybrid organic-inorganic halide perovskites are intensely studied due to their suitability for a wide range of optoelectronic applications, from solar cells with power conversion efficiencies exceeding 25\,\%~\cite{NREL}, to LEDs~\cite{Kumar2016,Yuan2016, Zhang2020} and radiation detectors~\cite{Stoumpos2013, Wei2016a}. Reducing the dimensionality of the inorganic sublattice by incorporation of large organic molecules such as butylammonium (BA) or phenethylammonium (PEA) affords bulk materials that combine many of the outstanding optoelectronic properties of the 3D parent materials, while exhibiting features of quantum and dielectric confinement~\cite{Katan2019, Hoye2022} and improved moisture stability~\cite{Smith2014, Quan2016}. Quasi-two-dimensional (2D) Ruddlesden-Popper (RP) perovskites with formula A$_{n-1}$A$_2$'B$_n$X$_{3n+1}$~\cite{Smith2014, Cao2015, Stoumpos2016, Gebhardt2017, Giri2020, Fridriksson2020} have shown promising performance as active materials in LEDs, lasers and solar cell devices~\cite{Tsai2016}. In these materials, A is an alkylammonium or arylalkylammonium cation separating the $n$-layer perovskite framework, which consists of a 2D network of corner-sharing BX$_6$ (B=Pb$^{+2}$, X=I$^-$, Br$^-$) octahedra with small cations A' (A'=CH$_3$NH$_3^+$, Cs$^+$) providing charge balance.

While the role of the dynamic distortions present in the 3D parent perovskite MAPbI3 has been intensively studied~\cite{Leguy2015b, Brivio2015a, Mayers2018, Gold-Parker2018, Ferreira2018, Jinnouchi:prl19, Gehrmann2019, Bischak2020, Bokdam:jpcc21}, the same cannot be said about the 2D derivatives, where the required time scale for accurately capturing and describing the change in the crystal structure was hitherto inaccessible due to significant computational requirements. The optical properties of RP perovskites are known to be governed by dielectric and quantum confinement, which has been shown experimentally~\cite{Ishihara1989, Muljarov1989,Gauthron2010, Tanaka2005} and by means of first-principles and semiempirical electronic structure calculations~\cite{Blancon2018, Cho2019, Palummo2021, Filip2022}. In contrast, much less is known about the effects of the organic interlayer and its coupling to the inorganic sublattice on the electronic and thermal properties of these materials. Ultraviolet (UV) and inverse photoemission spectroscopy combined with density functional theory (DFT) calculations have been reported for BA$_2$PbI$_4$ and BA$_2$PbBr$_4$, showing less band dispersion and a larger density of states (DOS) at the band edges than in their 3D counterparts~\cite{Silver2018}. DFT calculations by Gebhardt \textit{et al.}~\cite{Gebhardt2017} showed that the electronic structure of the layered bulk phase of PEA$_2$PbI$_4$ is almost unaffected by reducing the dimensionality to a monolayer, suggesting weak interactions between the molecular and perovskite sublattices along the stacking direction. The interaction between these sublattices was shown to be governed by steric effects in a study by Du \textit{et al.}~\cite{Du2017}. Furthermore, polar structural distortions leading to chirality or Rashba-/Dresselhaus effects have been achieved through targeted templating with organic molecules \cite{Jana2020, Jana2021,chakraborty_rational_2023}.

The role of the organic cations on phonon and exciton dynamics in quasi-2D RP perovskites has been studied in several experimental and theoretical studies. The coupling between thermal vibrations and optoelectronic properties in RP perovskites was studied by several authors, and shown to lead to phonon-assisted scattering, charge-carrier screening, and charge-carrier localization~\cite{kang_dynamic_2017, Cortecchia2017, neutzner_exciton-polaron_2018, cortecchia_structure-controlled_2018, Dahod2019, Dahod2020, Menahem2021, Quan:pnas2021, Ziegler2022}. Quan~\textit{et al.}~showed that the pronounced anharmonicity of BA$_2$PbI$_4$ induced by the organic spacers leads to enhanced phonon scattering rates and facilitates vibrational relaxation coupled to photogenerated excitations~\cite{Quan:pnas2021}. Furthermore, differential scanning calorimetry and X-ray diffraction measurements by Dahod~\textit{et al.} revealed a reversible phase transition associated with a partial melting of organic spacers in BA-based RP perovskites with higher inorganic layer thickness~\cite{Dahod2019}. While the dynamic distortions markedly influence the free carrier masses and drift-diffusion transport mechanisms, they do not affect the exciton binding energies and exciton diffusion in quasi-2D RP perovskites~\cite{Ziegler2022}. Thermal transport properties of quasi-2D RP perovskites have come into focus only recently, when several experimental and theoretical studies~\cite{Giri2020, Li:nanol21, Thakur:mh22} reported ultralow thermal conductivities in A$_2$PbX$_4$ (A=BA, PEA, X=I, Br) using thermal reflectance measurements combined with molecular dynamics simulations based on a classical force field. A theoretical study using a similar force field by Fridriksson \textit{et al.}~also showed that length and aromaticity of the interlayer molecule significantly affect the structural rigidity of the perovskite sublattice~\cite{Fridriksson2020}. Temperature-dependent Raman measurements on A$_2$PbBr$_4$ further highlighted the role of the A site molecule, with PEA leading to strongly anisotropic vibrational modes related to the orientation of the phenyl ring with respect to the perovskite sublattice~\cite{Dhanabalan2020}. Polarization-orientation Raman measurements by Menahem \textit{et al.}~showed the importance of anharmonic thermal distortions for the order-disorder phase transition of BA$_2$PbI$_4$ at 274\,K~\cite{Menahem2021}.

In this work we study how the structure and structural dynamics of A$_2$PbI$_4$ (A=BA, PEA) affect the thermal and electronic properties of these materials in comparison with the quintessential 3D perovskite CH$_3$NH$_3$PbI$_3$ (MAPbI$_3$). We find that the organic interlayer strongly affects thermal properties. In particular, the thermal conductivity of PEA$_2$PbI$_4$ is lower than that of BA$_2$PbI$_4$ despite the more strongly pronounced dynamic distortions in the BA layer in the latter. A complex interplay between the structural dynamics of the organic interlayer and their volume-phase relationship determines thermal transport in these layered materials. We provide an atomistic explanation for these findings by using temperature-dependent Raman spectroscopy and molecular dynamics (MD) simulations based on machine-learned force fields with DFT accuracy. Our MD simulations also enable us to identify the contributions of the organic and inorganic sublattices to the structural dynamics and provide a mechanistic explanation for the BA$_2$PbI$_4$ phase transition. We then systematically study the electronic structure of these materials. Our DFT calculations which we combine with ultraviolet photoelectron spectroscopy (UPS) show that the electronic valence band density of states of all three materials is governed by structural distortions of the perovskite sublattice - an indirect effect of the interaction between the two sublattices.

All experimental and computational techniques are described in Section~\ref{methods}. We start the discussion of our results by investigating the macroscopic thermal properties of the two RP perovskites A$_2$PbI$_4$ (A=BA, PEA), and those of MAPbI$_3$ in Section~\ref{sec:thermal}. In Section~\ref{sec:electronic}, we study the effect of the structural distortions on the electronic structure of these materials. Section~\ref{sec:summary} summarizes our conclusions.

\section{Experimental and Computational Methods}\label{methods}

\subsection{Sample Preparation}
The BA$_2$PbI$_4$ quasi-2D perovskite was prepared following the procedure described by Stoumpous~\textit{et al.} in Ref.~\citenum{Tsai2016}. The PEA$_2$PbI$_4$ samples were synthesized by our own procedure. A detailed description of the preparation method can be found in the Supporting Information (SI). The obtained (orange) BA$_2$PbI$_4$ and (yellow) PEA$_2$PbI$_4$ crystals and their corresponding powder X-ray diffraction (XRD) spectra are shown in Fig.~S1.

\subsection{Structures}
The RP perovskite BA$_2$PbI$_4$ is orthorhombic at room temperature (RT), with $Pbca$ symmetry. Furthermore, the perovskite layers consisting of halide octahedra with Pb centers are separated along the stacking direction by long organic butylammonium (BA) molecules, resulting in a longer out-of-plane lattice parameter $c = 26.23$\,\AA, and non-equal in-plane lattice parameters $a = 8.43$\,\AA~and $b = 8.99$\,\AA.

The replacement of the cation molecule in the previously described RP layered perovskite with phenethylammonium (PEA) leads to a more stable RP perovskite. PEA$_2$PbI$_4$ is triclinic at RT, with $P\bar{1}$ symmetry, rendering similar in-plane lattice parameters $a = b = 8.74$\,\AA~and an even larger out-of-plane lattice parameter $c = 32.99$\,\AA.

MAPbI$_3$ hybrid halide perovskite is known to occur in tetragonal phase, with $I4/mcm$ symmetry at RT, consisting of a 3D lattice of corner sharing PbI$_6$ octahedra, with methylammonium (MA) molecules in between. Thus, we compute the electronic properties of the RT $I4/mcm$ crystal structure, with lattice parameters $a = b = 8.97$\,\AA~and $c = 12.68$\,\AA, in good agreement with the previews reported values~\cite{Kawamura2002, Li2015d}. It is known that MAPbI$_3$ undergoes a phase transition at $T=327$\,K~\cite{Poglitsch1987}.

\begin{figure*}[ht]
\centering
  \includegraphics[width=\linewidth]{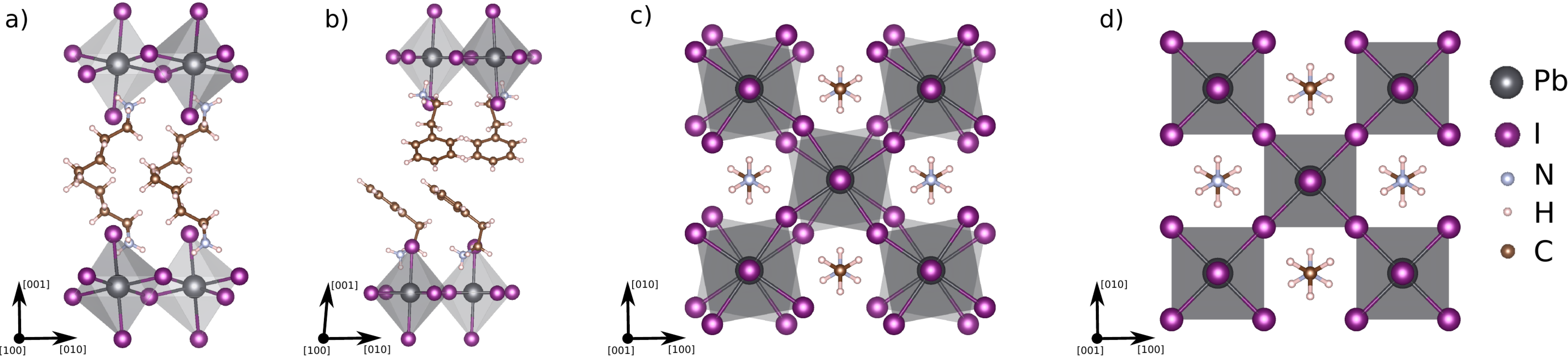}
  \caption{Schematic representation of the crystal structures used for all (static) DFT calculations. (a) $Pbca$ phase of BA$_2$PbI$_4$, (b) $P\bar{1}$ phase of PEA$_2$PbI$_4$, (c) $I4/mcm$ phase of MAPbI$_3$ and (d) $Pm\bar{3}m$ phase of MAPbI$_3$. In the two MAPbI$_3$ phases, the MA molecules were oriented with the C-N axis pointing along the [001] direction and antiparallel such that their dipole moments cancel out.}
  \label{fig:structures}
\end{figure*}

The crystal structures of BA$_2$PbI$_4$ and PEA$_2$PbI$_4$ extracted from single crystal X-ray diffraction experiments under ambient conditions~\cite{Billing2007,Du2017inorg} are represented in Fig.~\ref{fig:structures}(a) and (b), respectively, while the high-temperature cubic ($Pm\bar{3}m$) and RT tetragonal ($I4/mcm$) phases of MAPbI$_3$ are shown in Fig.~\ref{fig:structures}(c) and (d). All structure parameters are listed in Table~S1 in the SI.

\subsection{Thermal Characterization}
Thermogravimetric analysis was performed using a Netzsch PERSEUS STA 449 F3 Jupiter in platinum crucibles under nitrogen flow. A heating rate of 10\,K\,min$^{-1}$ was employed.

The heat capacity at constant pressure C$_p$ was determined by differential scanning calorimetry measurements according to the ASTM E1269 standard. The measurements were performed on a TA instruments Discovery DSC 2500. The temperature profile ranged from 193\,K to 453\,K using a heating rate 20\,K\,min$^{-1}$ with a nitrogen flow of 50\,mL\,min$^{-1}$. Two heating cycles were conducted and only the second cycle evaluated. Three samples for each material were prepared and averaged. Molar heat capacities were then calculated using the molar mass of the corresponding material. 

The density at RT was determined by helium pycnometry using an Ultrapyc 1200e (Quantachrome Instruments). Prior to each measurement the volume of the empty cell was calibrated. Afterwards, the samples were weighted into the cell and measured at 17\,psi of helium after the cell was purged with helium for one hour. The sample volume and density were related to the pressure drop when opening the valve to a second known volume. One hundred runs were performed and averaged to extract the density at RT. 

The temperature-dependent density was calculated by determining the thermal expansion coefficient. The change in volume of small sample fragments in the temperature regime between 283\,K and 383\,K was determined using an Olympus LEXT OLS5000 confocal microscope. The temperature was set with an Instec TS102G/V Peltier stage in steps of 20\,K. The 3D-scans were conducted using a $20\times$ objective with a numerical aperture of 0.45 and a working distance of 6.5\,mm. To detect the whole fragment several images were stitched together, and the volume was determined using the LEXT analysis application after performing a tilt correction on the substrates surface. Three sample fragments for each material were averaged.

Thermal diffusivity was determined by employing XFA (Xenon-Flash Analysis) measurements using a Netzsch LFA 467 HT. For this, cylindrical pellets of 10\,mm in diameter and approximately 1\,mm in height were pressed from each material and coated with graphite. It has to be noted that thereby effective thermal properties are determined, as opposed to single crystallite measurements predominantly done in literature~\cite{Rasel2020, Giri2020, Ge2018, Heiderhoff2017}. We, consequently, do not expect any anisotropic effects in these pellets. The thickness of the samples was determined using the Olympus LEXT OLS5000 confocal microscope, while the volumetric thermal expansion coefficient was assumed to be isotropic to correct for a change in sample thickness. Measurements were conducted under a nitrogen flow of 50\,mL\,min$^{-1}$ and each sample was measured three time at every temperature. The detectors field of view was improved using the build-in ZoomOptics system to only measure an area of 2.8\,mm in diameter. In this way a homogeneous sample thickness can be ensured while furthermore signal falsifications due to the sample environment (e.g. masks or apertures) are thus excluded. Data analysis was performed using a single-layer Cape-Lehman model taking into account heat losses on all sides.

\subsection{Raman Measurements}
The non-resonant Raman spectra were acquired using a micro-Raman spectrometer (WITec Alpha 300 RA+, Ulm, Germany) equipped with a UHTS 300 spectrometer and a back-illuminated Andor Newton 970 EMCCD camera. A frequency-doubled Nd-YAG laser with a wavelength of 785\,nm was used as the excitation source. The laser beam was focused onto the sample by means of a 50$\times$ long working distance (numerical aperture $\text{NA}=0.7$, lateral resolution ca. 500\,nm) Zeiss objective. The focal length of this spectrometer is 800\,mm, and it is equipped with a diffraction grating having a groove density of 1200 lines per millimeter resulting in a spectral resolution of $\sim$2\,cm$^{-1}$. The samples were excited with $\sim$10\,mW laser power in in-plane polarization mode and the spectra were collected after 50 accumulations with an integration time of 0.5\,s. All spectra were processed using a Witec Suite Five (version 5.2) software that allowed compensating for cosmic radiation and other background signals. For measurements, ca. 1\,$\mu{}$m thick BA$_2$PbI$_4$ and PEA$_2$PbI$_4$ pellets were prepared by pressing ca. 120\,mg of respective poly-crystalline grains at 5\,bar. 

\subsection{Computational Details}
To understand the electronic properties of the analysed materials, we carried out first principles density functional theory (DFT) calculations for both the experimental RT crystal structures and corresponding model systems. All DFT calculations were performed within the generalized gradient approximation of Perdew, Burke, and Ernzerhof (PBE)~\cite{Perdew1996} as implemented in the Vienna Ab$-$initio Software Package (\textsc{vasp})~\cite{Kresse1996a, Kresse1996b}, using projector augmented wave (PAW) pseudo-potentials~\cite{Kresse1999}, and including the effect of spin-orbit coupling~(SOC) self-consistently. The PAW pseudopotentials have the following electronic configurations: N ($2s^2 2p^3$), H ($1s^1$), C $2s^2 2p^2$, Pb ($6s^2 6p^2$) and I ($5s^2 5p^5$). For BA$_2$PbI$_4$ and PEA$_2$PbI$_4$ we sampled the first Brillouin zone using $\Gamma$-centered \textbf{k}-point grids with $2\times2\times1$ points for the ground-state calculations and $6\times6\times2$ points for the DOS calculations, respectively. For MAPbI$_3$ we employed a $6\times6\times6$ $\Gamma$-centered \textbf{k}-point grid. All calculations were performed using a cutoff energy for the plane-wave expansion of 500\,eV. To account for the on-average centrosymmetric structure of MAPbI$_3$ at RT and to avoid introducing artifacts related to the relative orientation of the MA molecules towards each other, we align the molecules such that they are anti-parallel and their net dipole moment is zero, as would be expected on average at RT~\cite{Leppert:jpcl2016}. The structural models that were used in our static DFT calculations and as starting point for our molecular dynamics simulations are depicted in Fig.~\ref{fig:structures}. The simulation of crystal structures using MD requires a potential energy surface (PES) with an accurate corrugation in the high dimensional parameter space of atomic coordinates. The PBE functional was shown to be a good approximation to the PES of MAPbI$_3$ through comparison with accurate calculations using the Random Phase Approximation (RPA)\cite{Bokdam:prl2017}.

Machine-learned force fields~(MLFF) were trained for BA$_2$PbI$_4$ and PEA$_2$PbI$_4$ during MD simulations with \textsc{vasp} v6.3, based on calculated total energies, forces and stress tensors for automatically (on-the-fly) selected structures in the isothermal–isobaric~($NPT$) ensemble. This approach is described in detail in Refs.~\citenum{Jinnouchi:prl19} and~\citenum{Jinnouchi:prb19}. The coefficients of the MLFF are re-optimized after every DFT step. A variant of the GAP-SOAP~\cite{Bartok:prl10,Bartok:prb13} method is
used as a descriptor of the local atomic configuration around each atom. Within a cutoff of 8\,\AA{} a two-body radial probability distribution $\rho^{(2)}_i$ is build, as well as a three-body angular distribution $\rho^{(3)}_i$ within a cutoff of 5\,\AA{}. A The atomic coordinates are smeared in the distributions by placing Gaussians with a width of 0.8\,\AA{}. The obtained distributions are projected on a finite basis set of spherical Bessel functions multiplied with spherical harmonics. The number of Bessel functions for both the radial and angular part is set to 8. The angular part has a maximal angular momentum of $l_{max}=4$. The coefficients of the projections are gathered in the descriptor vector $\mathbf{X}_i$. A kernel-based regression method is applied to map the descriptor to a local atomic energy. The similarity between two local configurations is calculated by a polynomial kernel function of power 4 containing a sum of scalar products of the two- and three-body descriptor vectors weighted by 0.1 and 0.9, respectively. 

Training started from the relaxed structures corresponding to those represented in Fig.~\ref{fig:structures} during 100\,ps at 400\,K with a timestep of 1.5\,fs. The PBE functional, a 300\,eV plane-wave cutoff, a $2\times2\times1$ $\Gamma$-centered \textbf{k}-point grid and a Langevin thermostat were applied. This was followed by constant temperature training at 300\,K and 450\,K, both for 100\,ps. Training was concluded with a cooling down from 350\,K to 150\,K in 100\,ps. In total 1245/1160 DFT calculations were performed for BA$_2$PbI$_4$/PEA$_2$PbI$_4$, respectively. From these 424, 1265, 603, 165, 2004 / 144, 1200, 859, 134, 1769 (Pb, I, C, N, H) local atomic configurations were selected and constitute the basis of the finished MLFFs. These force fields were used to calculate the vibrational density of states (VDOS) of 300\,K and 450\,K equilibrated $2\times2\times1$ supercells in the $NVE$ ensemble. The open source code: \href{http://www.dynamicsolids.net/code/dsleap}{Dynamic Solids Large Ensemble Analysis Package (DS-LEAP)} was used to calculate the VDOS from the 150\,ps long trajectories.

\subsection{UPS measurements}
UPS measurements were carried out on a PHI 5000 VersaProbe III system fitted with a He discharge light source providing stable and continuous He I and He II lines, under ultrahigh vacuum ($\sim$10$^{-10}$\,mbar). Clean ITO substrates ($14 \times 14$\,mm) pre-treated with O$_3$/UV for 15\,min at 323\,K were used as standard substrates for all UPS measurements. All samples were prepared via solution processing using anti-solvent method. For MAPbI$_3$, 1M PbI$_2$ and equimolar methyl-ammonium iodide (MAI) were dissolved in DMF. The solution was stirred at RT overnight, and filtered using a 0.2\,$\mu{}$m PTFE filter. The obtained clear solution was spin-coated onto a clean ITO; while spinning, after 8\,s from the start, 200\,$\mu{}$L of toluene was dripped over the substrate. Further, the substrate was annealed at 373\,K for 25\,min to obtain uniform MAPbI$_3$ film. PEA$_2$PbI$_4$ and BA$_2$PbI$_4$ were also prepared following the same procedure except that the stock solutions were prepared by dissolving crystals of respective materials in DMF. The samples were directly transported to the UPS instrument by using a N$_2$ filled, sealed stainless steel transport vessel without exposing them to the ambient conditions. All measurements reported in this study were carried out with the He I (21.22\,eV, 40\,W) line with -5\,V sample biasing and the corresponding photoemission with 90\,$^{\circ}$ take-off angle was collected at a multichannel analyzer. The Fermi level ($E_\text{F}$) and vacuum level ($E_{\text{vac}}$) were determined using a sputter cleaned gold foil. Work function and ionization potential values of the perovskites are calculated as the energy difference between $E_{\text{vac}}$ and $E_{\text{F}}$, and $E_{\text{vac}}$ and onset of the valence band, respectively. The resolution of the UPS measurements is $\pm 0.15$\,eV, calculated using the Fermi edge full-width-half-maximum of the gold spectrum and the presented work function and ionization potential values are reproducible within $\pm 0.05$\,eV, consistent with the resolution limit.

\section{Results and Discussion}

\subsection{Thermal Properties}\label{sec:thermal}
\begin{figure}[ht]
\centering
  \includegraphics[width=\linewidth]{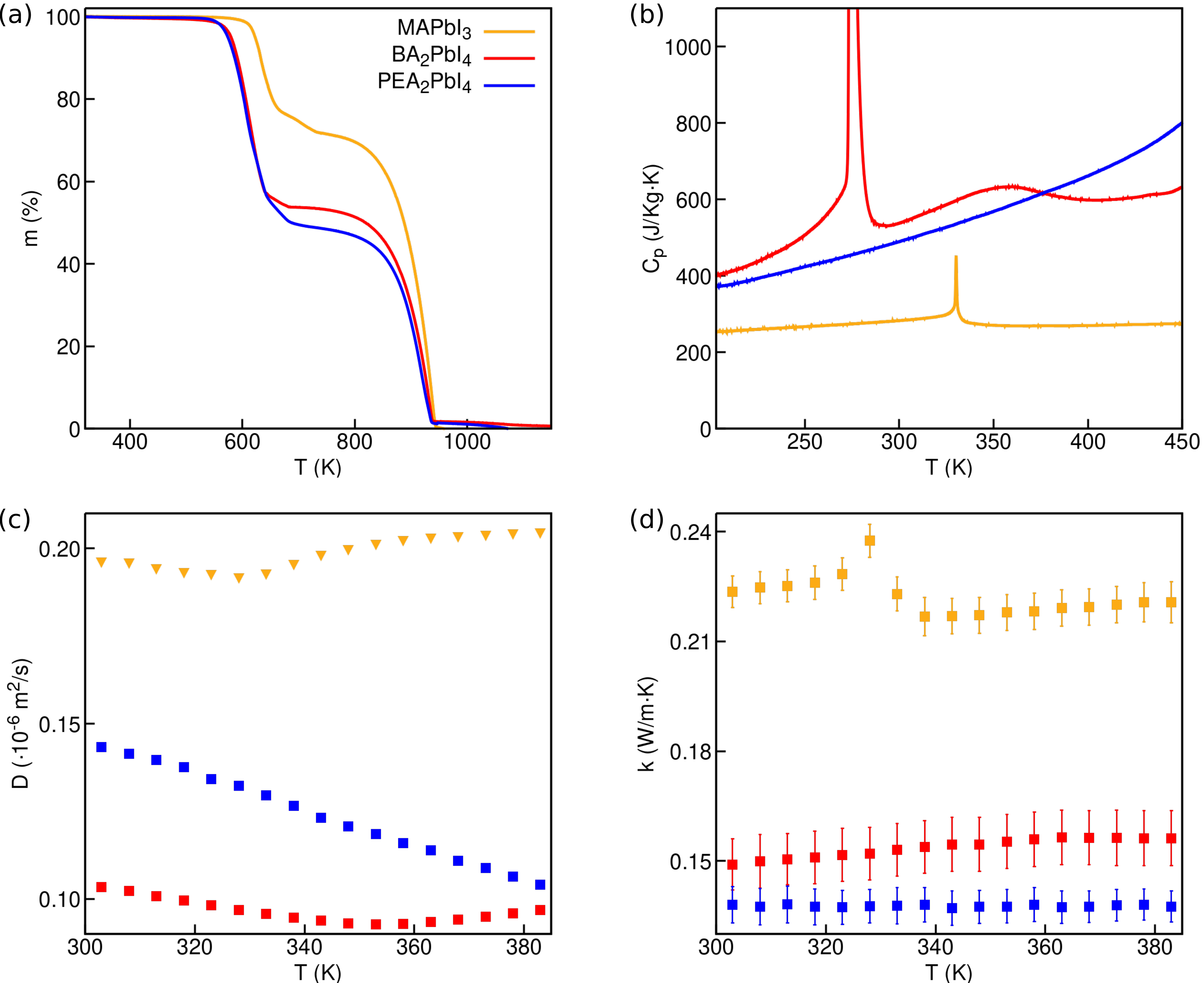}
  \caption{Thermal properties of MAPbI$_3$ (yellow), BA$_2$PbI$_4$ (red) and PEA$_2$PbI$_4$ (blue): (a) thermogravimetric analysis, (b) heat capacity, (c) thermal diffusivity and (d) determined thermal conductivity.}
  \label{fig:thermal}
\end{figure}

We start by verifying the thermal stability of all three compounds using thermogravimetric analysis shown in Fig.~\ref{fig:thermal}(a). In general, the two RP materials show a slightly earlier decomposition onset than MAPbI$_3$, with minor differences between PEA$_2$PbI$_4$ and BA$_2$PbI$_4$. All materials decompose step-wise: the organic constituents decompose first at $\sim$$573\,\text{K}$ -- $773\,\text{K}$, while the inorganic constituents decompose above $773\,\text{K}$. Note that both RP materials additionally exhibit light-induced degradation~\cite{Fang2018}.

Next, we determine the thermal conductivity using $k=D\cdot c_{\mathrm{v}}$, where we measured the thermal diffusivity $D$ (Fig.~\ref{fig:thermal}(c)) and approximated the volumetric heat capacity as the product of specific heat at constant pressure (Fig.~\ref{fig:thermal}(b)) and density: $c_{\mathrm{v}}=c_{\mathrm{p}} \cdot \rho$. Note, that in the following we assume the density to be constant (Table~S1), which is justified by the small temperature range we analyse~\cite{Ge2018, Heiderhoff2017}. Peaks in the specific heat in Fig.~\ref{fig:thermal}(b) are attributed to phase transitions. While MAPbI$_3$ transitions from a tetragonal to cubic symmetry at $327\,\text{K}$, BA$_2$PbI$_4$ has an orthorhombic structure throughout the entire analysed temperature range. The transition at $274\,\text{K}$ corresponds to a shift of the essentially rigid BA cations relative to the inorganic layers~\cite{Billing2007, Wang2016a, Rasel2020}: they are are tilted with respect to the inorganic layers below $274\,\text{K}$, but almost perpendicular above the phase transition temperature. This straightening of the organic molecules determines an increased interlayer spacing~\cite{Billing2007}. The heat capacity is a measure of how many vibrations can be excited inside the material. Therefore, it is markedly larger for BA$_2$PbI$_4$ and PEA$_2$PbI$_4$ as compared to MAPbI$_3$. Notably, the heat capacity of PEA$_2$PbI$_4$ is monotonically increasing, characteristic for an organic material, while it levels off for BA$_2$PbI$_4$ at approximately $373\,\text{K}$. 

Fig.~\ref{fig:thermal}(c) shows that PEA$_2$PbI$_4$ has a higher diffusivity than BA$_2$PbI$_4$, which may be explained by $\pi$-stacking of the phenyl rings facilitating transport between neighboring cations. Thermal diffusivity is a measure for the heat flux that is induced upon applying a temperature gradient. The different temperature dependencies of the thermal diffusivity are a measure of how fast a temperature change can equilibrate in the sample as a function of temperature. Here, MAPbI$_3$ and BA$_2$PbI$_4$ exhibit only a weak temperature dependency in the studied temperature regime, while for PEA$_2$PbI$_4$, a linear decrease with increasing temperature is apparent.

The thermal conductivities determined via $k=D\cdot c_{\mathrm{p}} \cdot \rho$ and showed in Fig.~\ref{fig:thermal}(d), also feature only minor temperature dependencies. In agreement with recent experimental and theoretical studies~\cite{Ge2018,Li:nanol21,Thakur:mh22}, we observe that both RP materials have lower thermal conductivities than MAPbI$_3$. Furthermore, the thermal conductivity of PEA$_2$PbI$_4$ is lower than that of BA$_2$PbI$_4$. These findings indicate that the thermal conductivity decreases with increasing organic layer thicknesses, or an increased amount of organic constituents, in agreement with a recent study on the chain-length dependence of thermal conductivity in extended alkylammonium chains with the general formula $(\mathrm{C}_{n}\mathrm{H}_{2n+1}\mathrm{NH}_{3})_2\mathrm{PbI}_4$~\cite{Rasel2020}. The drop in the thermal conductivity of the RP perovskites as compared to MAPbI$_3$ is expected since the organic-inorganic interface strongly hampers thermal transport in the out-of-plane direction. Such behavior is known for hybrid materials, e.g., nacre-mimetics~\cite{dorres_nanoscale-structured_2022}.

To probe the nature of the phase transitions observed in the thermal properties of Fig.~\ref{fig:thermal} and differentiate the underlying structural changes in BA$_2$PbI$_4$ and PEA$_2$PbI$_4$, we carried out molecular dynamics (MD) simulations of both materials at constant temperature and pressure, as well as under slow heating conditions (heating rate of $3\,\text{K/ps}$). We note that this finite heating rate is still fast as compared to experimental conditions and, as previously demonstrated in Ref.~\citenum{Bokdam:jpcc21}, leads to an overestimation of the phase transition temperature. The interatomic potentials were created using a state-of-the-art \textit{on-the-fly} machine learning approach~\cite{Jinnouchi:prl19}. This method has near DFT accuracy and the clear advantage that it also captures the anharmonic effects related to the conformational changes of the molecules at elevated temperatures~\cite{Bokdam:jpcc21,Lahnsteiner:prb22}. The resulting crystal dynamics at 300\,K are shown in supplementary movies 1 \&{} 2. For clarity, we removed the fast oscillations above 100\,cm$^{-1}$ from the trajectory by computing a running average with a window size of 333\,fs. The typical tilting and rotation pattern of the PbI$_6$ octahedra in the structures of Fig.~\ref{fig:structures}(a) and (b) are retained at 300\,K. The BA molecules allow for tilts of the octahedra away from the direction orthogonal to the perovskite plane, but show (on time-average) no rotations in this plane. The opposite occurs for PEA$_2$PbI$_4$, where no tilts, but alternating in-plane clockwise and counter-clockwise rotations of the octahedra are apparent. Contrary to MAPbI$_3$ where the ammonium group is able to attain different orientations depending on temperature, in BA$_2$PbI$_4$ and PEA$_2$PbI$_4$ the ammonium group is immobilized with respect to the Pb-I framework. On average, we find that in PEA$_2$PbI$_4$, the orientations of the organic moetities with respect to the Pb-I framework are similar to the tetragonal phase of MAPbI$_3$, whereas  in BA$_2$PbI$_4$ they are oriented similar to the orthorhombic phase~\cite{Lahnsteiner:prb16,Jinnouchi:prl19}. This reorientation of the molecules leads to equivalent distortions in the PbI layer that we denote as $a^0a^0c^+$ (tetragonal) and $a^-b^+a^-$ (orthorhombic) in Glazer notation~\cite{Whitfield:sr16}, for PEA$_2$PbI$_4$ and BA$_2$PbI$_4$, respectively.

The dynamics in the BA- and PEA-based perovskites also differ from MAPbI$_3$. MA molecules can rotate spherically in the cavities spanned by the Pb-I framework of MAPbI$_3$. In contrast, our MD simulations (starting at 250\,K and slowly heating up to 500\,K) show that the longer tails of the PEA and BA molecules serve as anchors, creating a large barrier for spherical rotation, but allowing for conical rotation with respect to the molecular backbone. As a result, the RP structures cannot reach cubic symmetry. The diverging (red) compared to the smooth (blue) line in Fig.~\ref{fig:thermal}(b) indicates that a much larger crystal lattice change is expected in the heating simulation of BA$_2$PbI$_4$ compared to PEA$_2$PbI$_4$.
\begin{figure}[ht]
\centering
  \includegraphics[width=0.80\linewidth]{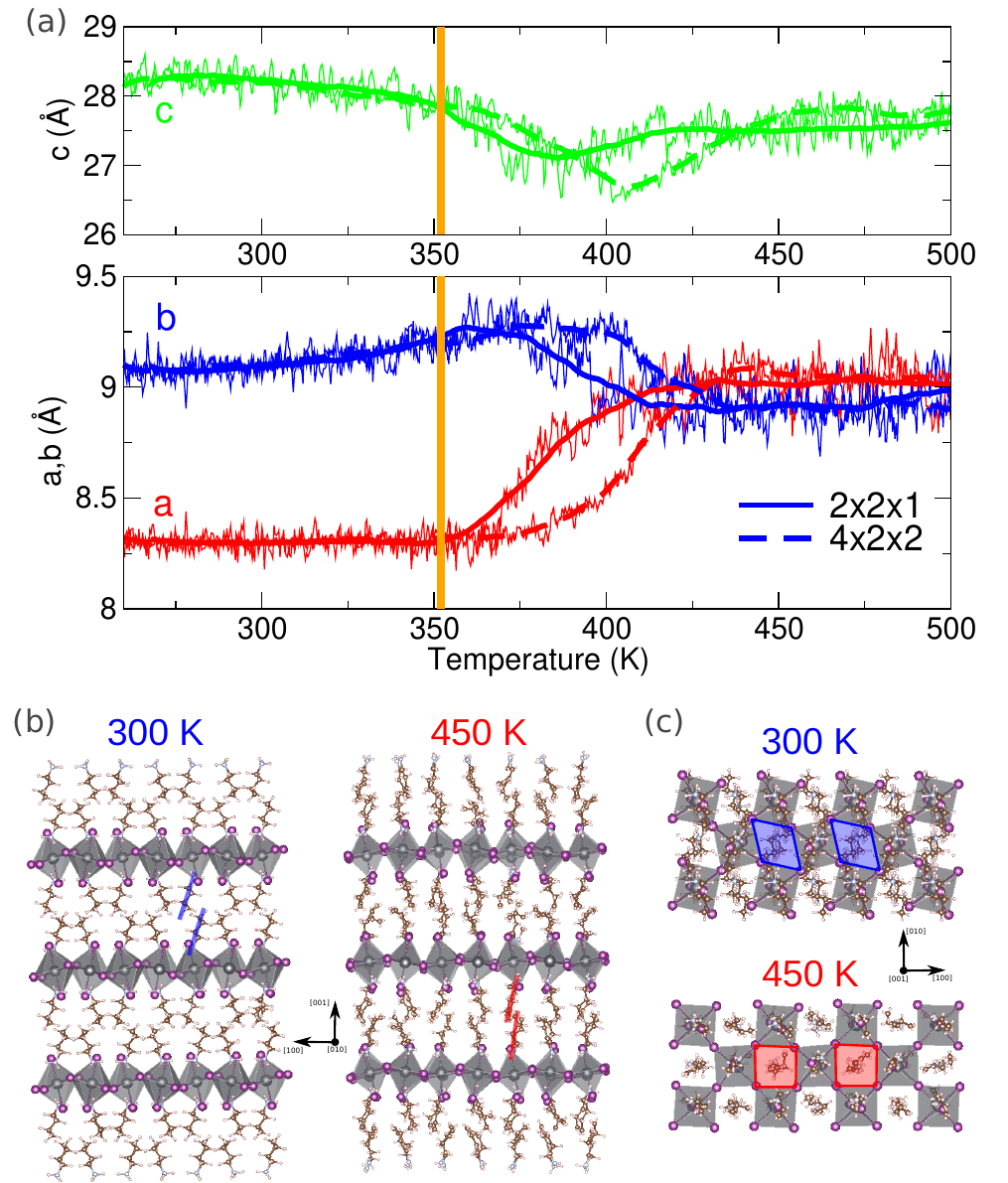}
  \caption{Lattice structure of BA$_2$PbI$_4$ as a function of temperature. (a) The out-of-plane $c$ (green) and in-plane $a$ (red), $b$ (blue) lattice constants characterizing an orthorhombic ($T<350$\,K) and tetragonal ($T>420$\,K) perovskite phase. The orange line marks the phase transition temperature $\rm T_c$. Side view (b) and top view (c) structures of the orthorhombic and tetragonal supercells obtained by averaging over snapshots during heating simulation of a $4\times2\times2$ supercell between 250-350\,K and 400-500\,K, respectively. The blue (red) solid line in the side view of panel (b) is a guide to the eye showing the orientation of BA molecules in the orthorhombic (tetragonal) phase. The cavities in which the BA molecules reside are highlighted in the top view of panel (c) by blue and red polygons, respectively.}
  \label{fig:phase-trans}
\end{figure}

The simulated phase transition of BA$_2$PbI$_4$ represented in Fig.~\ref{fig:phase-trans}(a) is a slow process, which does not happen instantaneously: it begins when the lattice constants start to change and ends when the lattice constants equilibrate. Since the simulations were performed under a fixed heating rate, we define the phase transition temperature (T$_c$) as the tipping point when the lattice geometry starts to modify, as indicated by the vertical orange line in Fig.~\ref{fig:phase-trans}(a). The same behavior was also observed for the 3D perovskites in Ref.~\citenum{Bokdam:jpcc21} and~\citenum{Fransson2023}. The transition temperature ($\rm T_c$) is overestimated as compared to the experimental $\rm T_c$ at $\sim274$\,K, but shows good qualitative agreement for the lattice parameters. A larger $4\times2\times2$ supercell shows that the simulated $\rm T_c$ value is converged with respect to supercell size. The transition to the high-temperature phase is slightly retarded, as a consequence of the finite heating rate in our simulations~\cite{Bokdam:jpcc21}. The crystal structures above and below $\rm T_c$ have two key differences as shown in Fig.~\ref{fig:phase-trans}(b) and (c), respectively:  $i)$ the packing of the molecules in the 2D crystals displays disorder at high temperature, and $ii)$ the  Pb-I framework in the $ab$-plane looses its alternating clockwise, anti-clockwise rotation pattern and becomes (on time-average) straight.
\begin{figure}[ht]
\centering
  \includegraphics[width=1.\linewidth]{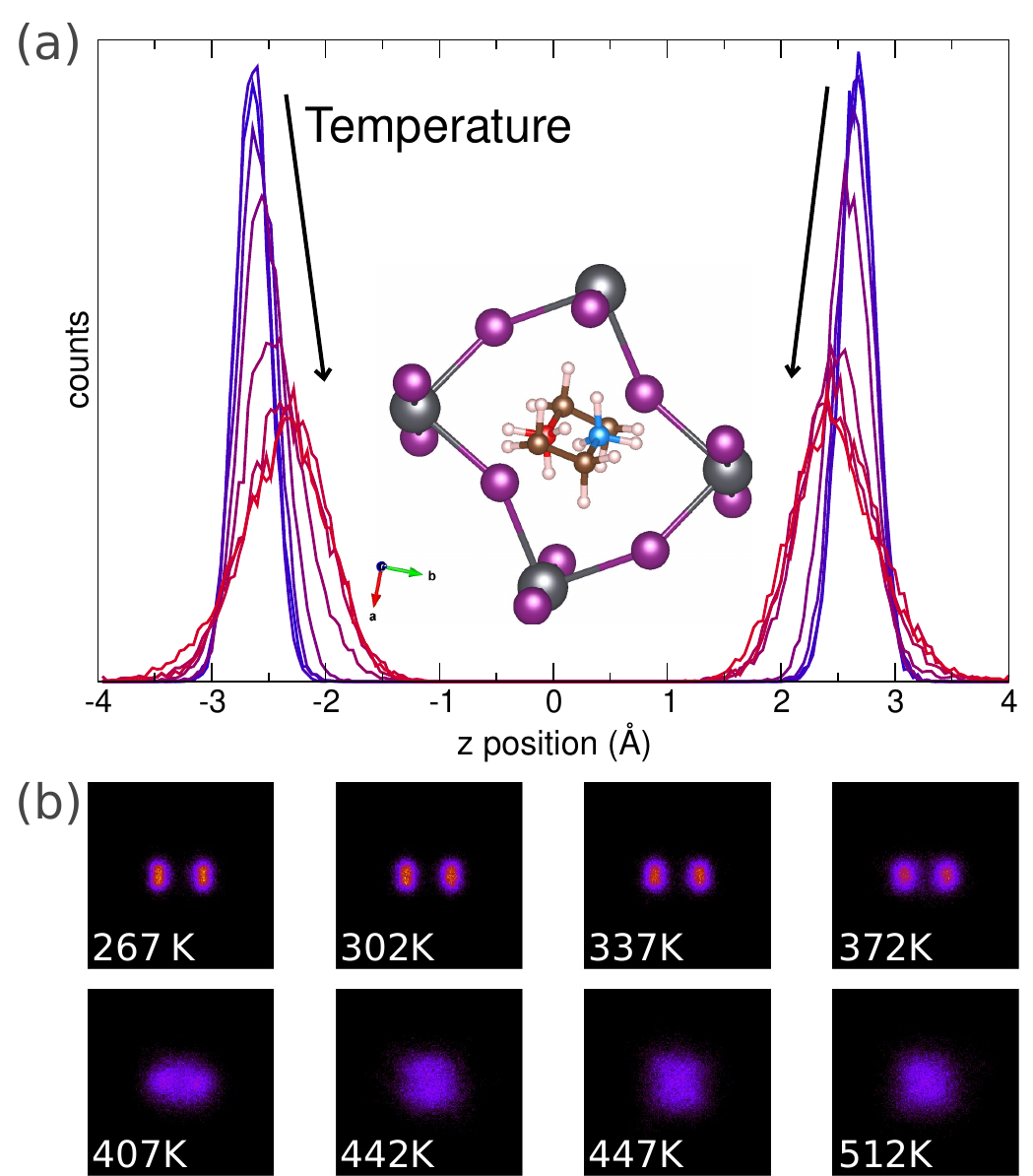}
  \caption{Phase transition mechanism. Temperature dependent distributions of the nitrogen penetration depth (a) and its in-Pb-plane position (b).}
  \label{fig:phase-trans-mechanism}
\end{figure}

Our MD trajectories further show that the phase transition mechanism is related to an increase of the ammonium group's \textit{penetration depth}~\cite{Li:cs19} and an activation of a \textit{rattling motion} above $\rm T_c$. In Fig.~\ref{fig:phase-trans-mechanism}(a) we show the temperature-dependent distribution functions of the nitrogen penetration depth defined as the orthogonal distance from the plane spanned by the Pb atoms to the nitrogen atom. We observe that the average moves into the direction of the Pb-plane as the temperature increases. Rattling of the ammonium group can be seen in the polar distributions of Fig.~\ref{fig:phase-trans-mechanism}(b), obtained by counting the $xy$ coordinates of the nitrogen atom in the Pb-I plane. Below $\rm T_c$ we find two fixed positions of nitrogen, corresponding to two distinct orientations of the ammonium group. When raising the temperature, rotations of the PbI$_6$ octahedra facilitate switching between these two orientations. After the transition is complete, the distribution displays a single maximum, in the center of the PbI$_6$ cage.

In agreement with the experimental observations shown in Fig.~\ref{fig:thermal}(b), our MD simulations of PEA$_2$PbI$_4$ do not display any abrupt changes in the lattice parameters within this temperature range. The conical rotation of the ammonium group of PEA that can occur at 400\,K is a rare event on the timescale accessible to our simulations and not a 'propeller' type motion. In agreement with our measured heat capacities, one can thus conclude that PEA$_2$PbI$_4$ exhibits a lower degree of dynamic distortions than BA$_2$PbI$_4$.

\subsection{Structural Dynamics}\label{sec:dynamics}
\begin{figure}[ht]
\centering
  \includegraphics[width=.8\linewidth]{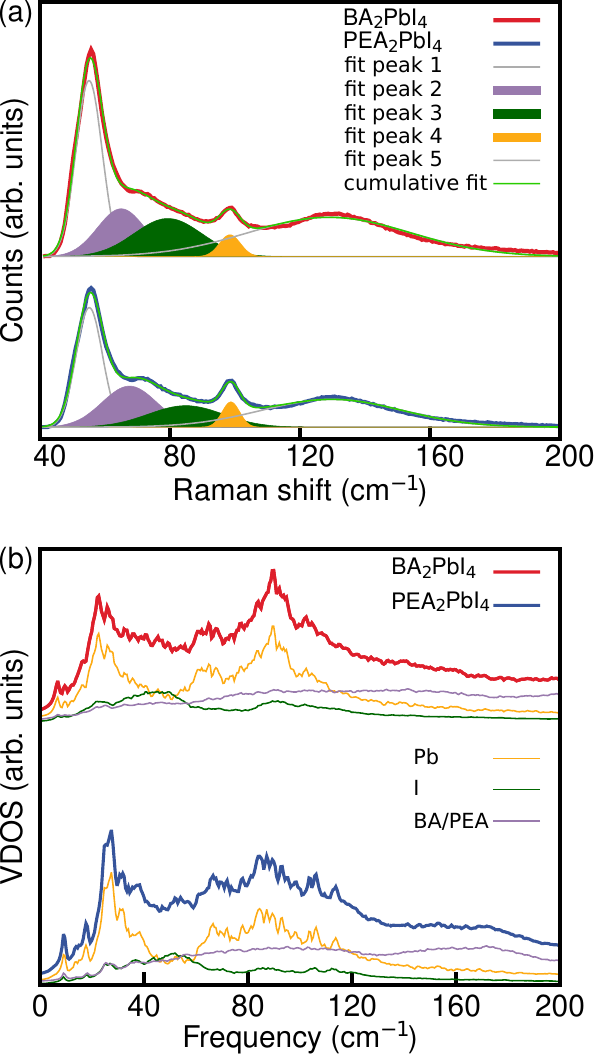}
  \caption{(a) Non-resonant Raman spectra of BA$_2$PbI$_4$ (red) and PEA$_2$PbI$_4$ (blue) pellets measured at RT (298 K) and the corresponding peak fits including the cumulative peak; the dash-dot lines are guide to eye, marking the fitted peaks maxima; (b) Vibrational density of states on the low frequency range of BA$_2$PbI$_4$ (red) and PEA$_2$PbI$_4$ (blue), simulated at 300\,K using molecular dynamics, decomposed into atomic contributions.}
  \label{fig:raman-vdos}
\end{figure}

To understand the vibronic contributions towards dynamical structural distortions in BA$_2$PbI$_4$ and PEA$_2$PbI$_4$, we carried out low-frequency (below 200\,cm$^{-1}$) non-resonant Raman measurements in the temperature range 298$-$363\,K. In Fig.~\ref{fig:raman-vdos}(a), Raman spectra of BA$_2$PbI$_4$ and PEA$_2$PbI$_4$ pellets measured between 40 to 200\,cm$^{-1}$ at 298\,K are shown. 

The deconvolution and unambiguous spectral assignment of Raman peaks of quasi-2D perovskites is not straightforward and there is no consensus among earlier reports on similar systems. For example, Dragomir~\textit{et al.}~\cite{Dragomir2018}~investigated both BA$_2$PbI$_4$ and PEA$_2$PbI$_4$ and assigned all peaks observed below 100\,cm$^{-1}$ exclusively to in- and out-of-phase Pb-I stretching vibrations and octahedral rotations; spectral features in the 100$-$200\,cm$^{-1}$ range are attributed to convoluted torsional and libration modes of the organic cations, which is supported by a few other reports~\cite{Cortecchia2017, Ivanovska2016, Quarti2014}. On the other hand, Dhanabalan~\textit{et al.}\cite{Dhanabalan2020} and Maczka~\textit{et al.}~\cite{Maczka2019} identified the peaks obtained from similar quasi-2D perovskites located up to 150\,cm$^{-1}$ as originating from Pb-X stretching modes. In contrast, a detailed analysis of vibrational modes in tetragonal MAPbI$_3$ by Leguy \textit{et al.}~\cite{Leguy2016} attribute many Raman modes observed below 150\,cm$^{-1}$ to organic cation coupled lattice motions. 

Based on these reports, we deconvoluted both spectra measured at RT into five individual components (peak 1, 2, 3, 4 and 5) as shown in Fig.~\ref{fig:raman-vdos}(a)~\cite{Dragomir2018}. We note that, due to the cut-off frequency of the optical filter used, spectral features below 50\,cm$^{-1}$ are not considered in the discussion. However, our Raman spectra are in good agreement with the ones reported by Menahem~\textit{et al.}~\cite{Menahem2021} (see Fig.~S2). We obtained satisfactory fits by using Gaussian functions, indicating inhomogeneous broadening of the peaks arising from orientation and/or positional disorder of the organic cations in the Pb-I cages~\cite{Leguy2016} (see SI and Fig.~S3 for a more detailed discussion).

To explore the nature of these peaks, we carried out molecular dynamics (MD) simulations of both crystals at 300\,K. We simulated the vibrational density of states (VDOS) by calculating the (mass-weighted) velocity auto-correlation functions from 150\,ps long MD trajectories in the micro-canonical ensemble. The resulting total VDOS of both perovskites over a large frequency domain is shown in Fig.~S4. Above 200\,cm$^{-1}$ the spectra are quite different, demonstrating the characteristic intra-molecular vibrations of the BA and PEA molecules. However, below 200\,cm$^{-1}$ the spectra are remarkably similar, in agreement with our Raman spectroscopy experiments. Note that the Raman spectrum only shows dipole active modes, whereas the VDOS shows all atomic vibrations. In Fig.~\ref{fig:raman-vdos}(b) we have decomposed the low-frequency part of the VDOS into its atomic contributions. We observe that the two pronounced peaks at $\sim$25\,cm$^{-1}$ and $\sim$88\,cm$^{-1}$ are vibrations of the Pb ions. Since they are also visible in the iodine spectrum at the same frequencies, we can attribute these peaks to Pb-I modes. Both iodine spectra show a dominant peak at $\sim$48\,cm$^{-1}$, however a matching peak is not observed in the Pb spectrum. The organic components show a broad (relatively featureless) distribution in the low frequency spectrum, indicative of weak coupling to the Pb-I framework. Furthermore, our MD simulations explain the change of the intensity in the low frequency region observed in Ref.~\citenum{Menahem2021}: We propose that the observed increase in the intensity of the Raman spectrum around $\sim$25\,cm$^{-1}$ is a consequence of the low-frequency rattling motion of the ammonium group depicted in Fig.~\ref{fig:phase-trans-mechanism}(b) that becomes active at temperatures higher than $\rm T_c$ (see Fig.~S5 for details).

In summary, we show that the phase transition of BA$_2$PbI$_4$ at $\sim 274\,\text{K}$ observed in our thermal measurements is linked to the increase of the ammonium group penetration depth at elevated temperature. Dynamic distortions play a crucial role in determining the thermal properties of this quasi-2D perovskite.

\subsection{Electronic Structure}\label{sec:electronic}
\begin{figure*}[ht]
\centering
  \includegraphics[width=1.\linewidth]{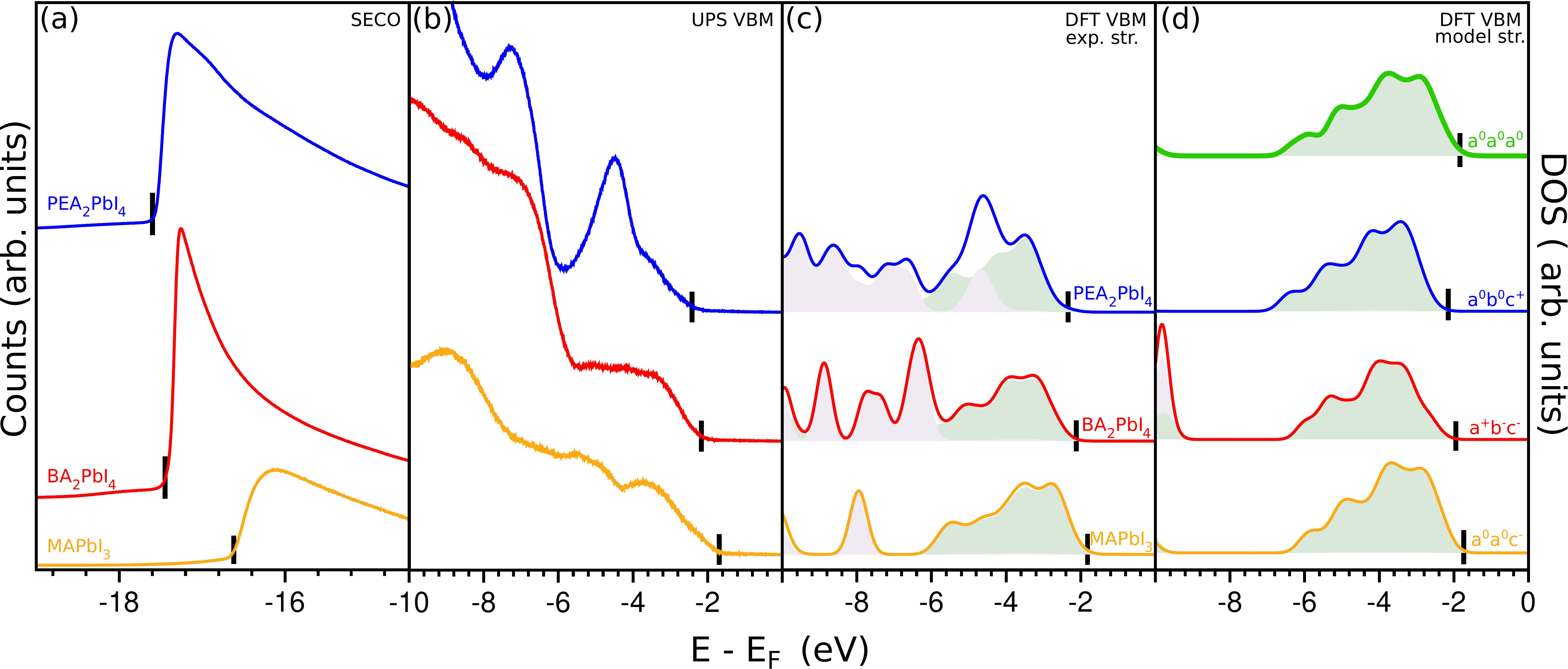}
  \caption{Ultraviolet photoelectron spectra (UPS) showing (a) the secondary electron cut-off (SECO) and (b) valence band maximum (VBM) regions. DFT-PBE total DOS for (c) experimental and (d) model structures of tetragonal MAPbI$_3$ (yellow), BA$_2$PbI$_4$ (red), PEA$_2$PbI$_4$ (blue) and cubic MAPbI$_3$ (green). All spectra are scaled with respect to the Fermi energy level and the black dash line marks the valence band onset. The shaded areas in the computed DOS represent the projections onto molecular orbitals with states derived from the inorganic and organic sublattices in green and purple, respectively.}
  \label{fig:UPS-DOS}
\end{figure*}

Our Raman measurements and MD simulations are indicative of a subtle interplay between the inorganic and organic sublattice in both BA$_2$PbI$_4$ and PEA$_2$PbI$_4$. Therefore, we further investigate how these effects influence the electronic structure of both materials, in comparison with the quintessential 3D perovskite MAPbI$_3$. To this end, we start by comparing the electronic density of states (DOS) of all three materials at room-temperature. Fig.~\ref{fig:UPS-DOS} shows the secondary electron cut-off (SECO) and valence band maximum (VBM) regions of BA$_2$PbI$_4$, PEA$_2$PbI$_4$, and MAPbI$_3$ obtained from ultraviolet photoelectron spectroscopy (UPS), as well as the first-principles DOS. The experimentally determined work function and ionization energies of the three perovskites are reported in Table~\ref{tab:work-function-ionization-potential} and are in close agreement with previous reports~\cite{Olthof2017, Silver2018, Yang2020a}.
\begin{table}[ht]
\small
  \caption{Experimentally determined work function and ionization energy (in eV) of MAPbI$_3$, BA$_2$PbI$_4$ and PEA$_2$PbI$_4$.}
  \label{tab:work-function-ionization-potential}
  \begin{tabular*}{\linewidth}{ccc}
    \centering
    \multirow{2}{*}{\textbf{System}} & \textbf{Work} & \textbf{Ionization} \\
    & \textbf{function (eV)} & \textbf{energy (eV)} \\
    \hline
    MAPbI$_3$      & $4.40$ & $6.09$ \\
    BA$_2$PbI$_4$  & $3.91$ & $6.08$ \\
    PEA$_2$PbI$_4$ & $4.68$ & $6.10$ \\ \hline
    \hline
  \end{tabular*}
\end{table}

The comparison of the DOS from standard semilocal approximations of DFT with experimental photoelectron spectra is known to be complicated by several factors. First, the eigenvalue corresponding to the VBM which should be equal to minus the ionization potential in exact Kohn-Sham DFT, is typically overestimated by several electronvolt by standard semilocal approximations like PBE~\cite{Kummel2008a}. We therefore red shift our calculated DOS such that it  matches the experimental VBM, i.e., the difference between the measured ionization energy and work function (see Table~\ref{tab:work-function-ionization-potential}). Second, energetically high-lying valence bands have been shown to be good approximations to the electron removal energies measured in a photoemission experiment, but only when calculated from very accurate electron densities~\cite{Chong2002}. Empirically, one finds that the DOS calculated with PBE is compressed with respect to experiment~\cite{Korzdorfer2010a,Egger2016}. We therefore follow the approach by Tao~\textit{et al.}~\cite{Tao:natcomm2019} and stretch our calculated DOS to match the measured spectra (corresponding to a stretch of $\sim$~2.6\,\%). A more rigorous approach would be the calculation of the electronic structure of all three materials within the GW approach which allows for a more straightforward comparison with experimental photoemission spectra but is outside the scope of this work~\cite{Filip2014d, Leppert:prm2019}. However, perfect agreement with experiment is not required for the peak assignment to the organic and the inorganic sublattices that we attempt in the following. 

Panel (c) of Fig.~\ref{fig:UPS-DOS} shows the contributions of the inorganic and organic sublattices to the total DOS in green and purple, respectively. For all three materials, the DOS in the vicinity of the VBM is derived from electronic states associated with Pb $s$ and I $p$ states (see Fig.~S6 and Fig.~S7 for complete band structures with orbital contributions). For MAPbI$_3$ we find, in agreement with previous work~\cite{Umari2014, Egger2016, Caputo2019}, that MA-derived electronic states lie far below the VBM, at $\sim$-8\,eV in our calculations. These states are well separated from the PbI-derived states at higher energies. This is different for the two RP compounds. While the DOS within $\sim$2\,eV below the VBM is dominated by Pb $s$- and I $p$-derived states for all three materials, the organic sublattice contributes already at $\sim$-4\,eV for PEA$_2$PbI$_4$ and $\sim$-6\,eV for BA$_2$PbI$_4$. In particular, the pronounced peak in the experimental spectrum of PEA$_2$PbI$_4$ at $\sim$-4\,eV arises partially from electronic states associated with the organic sublattice. We explain the differences between the three materials by computing the HOMO-LUMO gaps of the isolated molecular cations using DFT-PBE as implemented in the \textsc{turbomole} program package~\cite{Ahlrichs1989, Balasubramani2020}. In Table~\ref{tab:HOMO-LUMO-gap}, we compare these HOMO-LUMO gaps with the energy gap between the lowest conduction band and the highest valence band derived from the organic sublattice in the band structure of BA$_2$PbI$_4$, PEA$_2$PbI$_4$ and MAPbI$_3$ (represented in Fig.~S6).
\begin{table}[ht]
\small
  \caption{Calculated HOMO and LUMO eigenvalues and HOMO-LUMO gaps (in eV) of organic cations MA$^+$, BA$^+$ and PEA$^+$, energy gap (in eV) between the electronic state derived from the organic sublattice in the band structure of MAPbI$_3$, BA$_2$PbI$_4$ and PEA$_2$PbI$_4$.}
  \label{tab:HOMO-LUMO-gap}
  \begin{tabular*}{\linewidth}{lllll}
    \hline
    \textbf{} & \hfill\textbf{MA$^+$} & \hfill\textbf{BA$^+$} & \hfill\textbf{PEA$^+$}\\
    \hline
    HOMO & \hfill $-15.74$ & \hfill $-11.56$ & \hfill $-9.94$ \\
    LUMO & \hfill $-6.29$ & \hfill $-5.79$ & \hfill $-5.37$ \\
    HOMO-LUMO gap & \hfill $9.45$ & \hfill $5.77$ & \hfill $4.57$ \\
    Energy gap & \hfill $8.77$ & \hfill $7.50$ & \hfill $4.47$ \\ \hline
    \hline
  \end{tabular*}
\end{table}
For all three A site cations, the HOMO-LUMO gap is similar to the energy gap of the organic sublattice in the band structure.

Despite the significant differences in the electronic DOS related to the different energies of the organic sublattices, the DOS of the inorganic sublattice is similar for all three materials. This observation is also supported by the experimentally-determined ionization energies. As shown in Table~\ref{tab:work-function-ionization-potential}, the ionization energies of all three materials are within 20\,meV of each other, well within the resolution limit. However, our DFT calculations allow us to identify the atomistic origin of more subtle differences. We find that in the case of MAPbI$_3$, the DOS close to the VBM is primarily derived from the equatorial halides, i.e., the halide ions in the perovskite plane determined by the [100] and [010] directions (defined in Fig.~\ref{fig:structures}), whereas in the case of BA$_2$PbI$_4$ and PEA$_2$PbI$_4$ it is derived from contributions of the out-of-plane axial halides.

To assess the effect of the structural distortions on the electronic DOS of the quasi-2D perovskites, we constructed four model systems with layer thickness of $n=1$, by replacing the organic cation with Cs$^+$. The undistorted model system was constructed using untilted and undistorted metal-halide octahedra with $Pb-I$ bond lengths of 3.15\,\AA, equivalent to the cubic $Pm\bar{3}m$ phase of MAPbI$_3$. The three distorted model systems feature the same tilts and distortions as the experimental structures. We use Glazer notation to distinguish these four structures: undistorted model system -- $a^0a^0a^0$, MA-like distorted system -- $a^0a^0c^-$, BA-like distorted system -- $a^+b^-c^-$, PEA-like distorted system -- $a^0b^0c^+$. In order to avoid spurious interactions between periodic images in the model structures, a vacuum layer of at least 20\,\AA{} was introduced and a dipole correction was applied in all calculations. The electronic DOS of the model systems, shown in Fig.~\ref{fig:UPS-DOS}(d), is in close agreement with that of the experimental crystal structures, shown in Fig.~\ref{fig:UPS-DOS}(c). The comparison demonstrates that octahedral tilts and distortions that are a consequence of interactions at the organic-inorganic interface as shown in Section~\ref{sec:thermal}, govern the electronic DOS close to the VBM.

\begin{figure}[h!]
\centering
  \includegraphics[width=1.\linewidth]{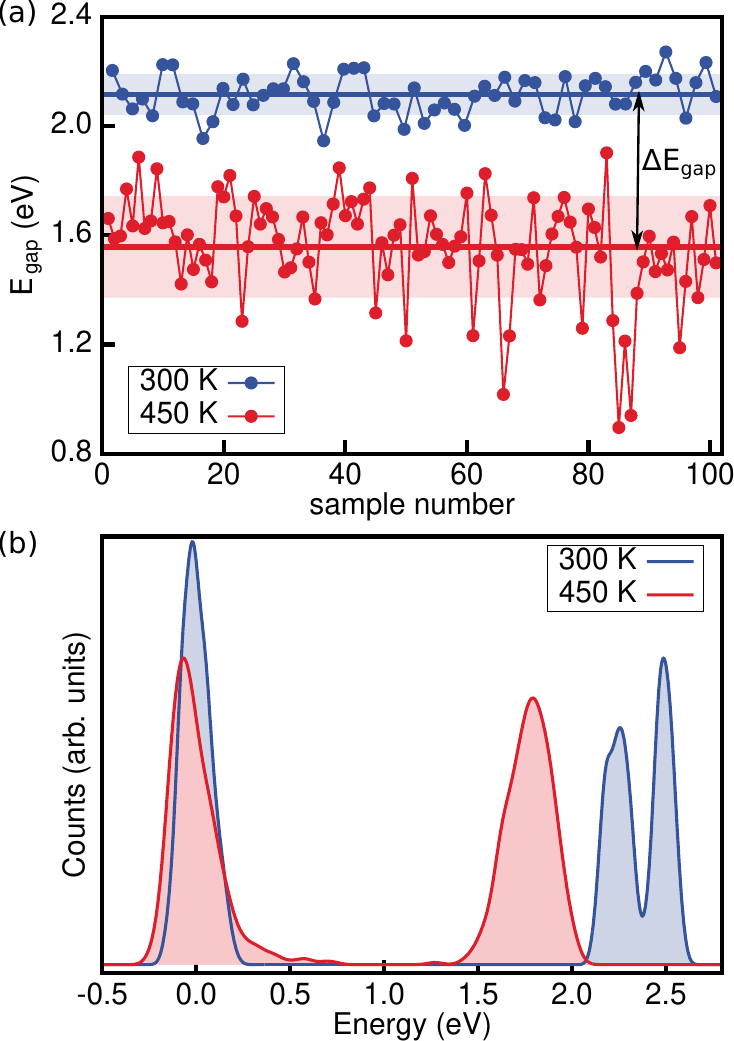}
  \caption{(a) Fundamental band gap of MD snapshots in the micro-canonical ensemble at 300\,K (blue) and 450\,K (red), respectively. The horizontal solid blue (red) line represents the average fundamental gap at 300\,K (450\,K). The variation of the average fundamental band gap $\Delta E_\text{gap}$ induced by the phase transition is represented by a black arrow. (b) DOS of the band edges at 300\,K (blue) and 450\,K (red). The energy axis is shifted such that zero of energy corresponds to the mean of the VBM region.}
  \label{fig:gap-fluct}
\end{figure}

Finally, we probe the effect of temperature and the phase transition discussed in Section~\ref{sec:thermal} on the electronic DOS of BA$_2$PbI$_4$. We start by comparing the fundamental band gap at the $\Gamma$-point below and above the phase transition temperature $\rm T_c$, respectively. We select 100 decorrelated snapshots ($2\times2\times1$ supercells) from the MD trajectories in the micro-canonical ensemble at 300\,K and 450\,K, respectively and compute their PBE (without SOC) band gap. By averaging over the band gaps of the selected structures at constant temperature, we estimate the fundamental band gap below ($E_\text{gap}^\text{300\,K}$) and above ($E_\text{gap}^\text{450\,K}$) $\rm T_c$. In Fig.~\ref{fig:gap-fluct}(a) we show the average fundamental band gap at finite temperature and its fluctuations. While the standard deviation of the band gap fluctuations at 300\,K is only 73\,meV, in agreement with previous work \cite{Ghosh2020:JMCA}, the gap fluctuations in the high temperature regime are twice as high. This wider spread is a consequence of larger distortions induced by the increased degree of disorder in the packing of molecules at 450\,K. We then compute the variation of the average fundamental band gap induced by the phase transition: $\displaystyle{\Delta E_\text{gap} = \abs{E_\text{gap}^\text{450\,K}-E_\text{gap}^\text{300\,K}}}$ and find that it closes by $\sim0.56$\,eV as a result of the phase transition.

To explain this change induced by the "straightening" of the Pb-I framework, we investigate the DOS of the band edges at 300\,K and 450\,K respectively. We obtain the DOS at elevated temperatures by averaging over the energies of the valence and conduction band edge states of all 100 structure snapshots, convoluted with a broadening function. For the broadening we use a Gaussian distribution $\displaystyle {g(x) = \frac {1}{\sigma}e^{-{\frac {1}{2}}\left({\frac {x-\mu }{\sigma }}\right)^{2}}}$ centered at each energy $\mu = E_i$ and with the standard deviation equal to the thermal energy $\sigma = k_B \cdot T$, where $k_B$ is the Boltzmann constant and $T$ is the temperature. As shown in Fig.~\ref{fig:gap-fluct}(b), the CBM DOS at 300\,K is split into two peaks. This is a consequence of the localization of the CBM charge density on the two inequivalent Pb-I layers of the Ruddlesden-Popper structure (see Fig.~S8). Each sub-peak comprises two states, with slightly different energies, localized in different inorganic layers. This feature is lost at 450\,K when the structure becomes more cubic and the energy differences between the conduction band states become negligible. Additionally, Fig.~\ref{fig:gap-fluct}(b) shows that the closing of the band gap above $\rm T_c$ is a result of the interplay of the inhomogeneous broadening of the VBM, which leads to the appearance of edge states "leaking" into the band gap and a red-shift of the CBM.

\section{Conclusions}\label{sec:summary}
In summary, we studied the thermal and electronic properties of the quasi-2D perovskites BA$_2$PbI$_4$ and PEA$_2$PbI$_4$ in comparison with MAPbI$_3$ first-principles calculations. Thermal characterization and Raman measurements combined with first-principles molecular dynamics simulations show a high degree of dynamic disorder in all three materials. However, we demonstrated that the coupled dynamics of their organic and inorganic sublattices differ considerably, leading to pronounced temperature-dependent differences in their thermal properties. In particular, we elucidated the mechanism for the 274\,K phase transition observed in BA$_2$PbI$_4$ and showed explicitly that PEA$_2$PbI$_4$ exhibits a lower degree of dynamic distortions than BA$_2$PbI$_4$. The average crystal structure arises as a consequence of these coupled dynamic distortions and has a significant effect on the electronic DOS. Our photoelectron spectroscopy and first-principles DFT calculations showed that the organic sublattice contributes electronic states at energies between 4\,eV and 8\,eV below the valence band maximum, depending on the organic cation. Furthermore, we explicitly showed that the details of the electronic DOS close to the valence band maximum are determined by the average structural distortions of the inorganic sublattice. Overall, our study demonstrates that the organic sublattice has pronounced effects on thermal and electronic properties. An atomistic understanding of these effects can thus be used to design new materials with tailored properties.

\begin{acknowledgement}
This work was supported by the Bavarian State Ministry of Science and the Arts through the Collaborative Research Network Solar Technologies go Hybrid (SolTech), the Elite Network Bavaria, and the German Research Foundation (DFG) through SFB840 B7 and Th 807/8-1, and through computational resources provided by the Bavarian Polymer Institute (BPI). This work was also supported by NWO Domain Science for the use of the supercomputing facilities. R.B. acknowledges support by the DFG program GRK1640.
\end{acknowledgement}

\begin{suppinfo}
Detailed description of sample preparation procedure\\
Additional discussion of Raman spectra\\
Additional theoretical results explaining the phase transition mechanism and impact of distortions on the electronic structure
\end{suppinfo}

\providecommand{\latin}[1]{#1}
\makeatletter
\providecommand{\doi}
  {\begingroup\let\do\@makeother\dospecials
  \catcode`\{=1 \catcode`\}=2 \doi@aux}
\providecommand{\doi@aux}[1]{\endgroup\texttt{#1}}
\makeatother
\providecommand*\mcitethebibliography{\thebibliography}
\csname @ifundefined\endcsname{endmcitethebibliography}
  {\let\endmcitethebibliography\endthebibliography}{}

\end{document}